\documentclass[conference]{IEEEtran}

\usepackage[noadjust]{cite}

\ifCLASSINFOpdf
   \usepackage[pdftex]{graphicx}
\else
   \usepackage[dvips]{graphicx}
\fi

\usepackage[cmex10]{amsmath}

\usepackage{algorithm}
\usepackage{algorithmic}

\usepackage{array}

\usepackage{mdwmath}
\usepackage{mdwtab}

\usepackage{eqparbox}

\usepackage[tight,footnotesize]{subfigure}

\usepackage{url}

\usepackage[font=footnotesize, justification=centering]{caption}

\begin{document}
\bstctlcite{IEEEexample:BSTcontrol}

\title{Design of a Novel Network Architecture for Distributed Event-Based Systems\\  Using Directional Random Walks in an Ubiquitous Sensing Scenario}
\author{\IEEEauthorblockN{Cristina Mu\~{n}oz and Pierre Leone}
\IEEEauthorblockA{Department of Computer Science\\
University of Geneva\\
Carouge, Switzerland\\
Email: \{Cristina.Munoz, Pierre.Leone\}@unige.ch}}

\maketitle

\begin{abstract}
Ubiquitous sensing devices frequently disseminate data among them. The use of a distributed event-based system that decouples publishers from subscribers arises as an ideal candidate to implement the dissemination process. In this paper, we present a network architecture that merges the network and overlay layers of typical structured event-based systems. Directional random walks are used for the construction of this merged layer. Our strategy avoids using a specific network protocol that provides point-to-point communication. This implies that the topology of the network is not maintained, so that nodes not involved in the system are able to save energy and computing resources. We evaluate the performance of the overlay layer using directional random walks and pure random walks for its construction. Our results show that directional random walks are more efficient because: (1) they use less nodes of the network for the establishment of the active path of the overlay layer and (2) they have a more reliable performance. Furthermore, as the number of nodes in the network increases, so do the number of nodes in the active path of the overlay layer for the same number of publishers and subscribers. Finally, we discard any correlation between the number of nodes that form the overlay layer and the maximum Euclidean distance traversed by the walkers.
\end{abstract}

\vspace{1em}
\begin{IEEEkeywords}
Distributed Event-Based Systems; Overlay Layer; Directional Random Walks; Pure Random Walks; Wireless Sensor Networks.
\end{IEEEkeywords}

\section{Introduction}
\label{sec:introduction}
Ubiquitous or pervasive computing \cite{FUTURECOMPUTING14DRW}\cite{Cook:2012:RPC:2109687.2109848} uses many sources and destinations to gather and process data related to physical processes with the aim of making possible human-computer interaction. In the process of dissemination, some devices generate the data, while others are waiting for the sensing data. In this context, the use of a distributed event-based system \cite{Muhl:2006:DES:1162246} arises as an ideal candidate to implement the model of communication on the reception or transmission of events. 

\begin{figure}[!t]
\centering
\includegraphics[width=2.75in]{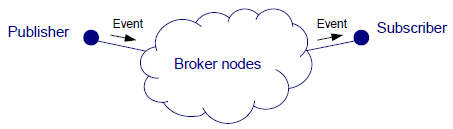}
\caption{Distributed notification service using a network of brokers.}
\vspace{-2.25em}
\label{PubSub}
\end{figure}

The main characteristic of an event-based system is that publishers and subscribers are decoupled. This means that they do not have any information about each other. The element in charge of matching notifications with subscriptions is called the event notification service. In distributed networks, the event notification service is implemented using a network of brokers nodes (see Figure~\ref{PubSub}). It is considered that a broker is any node in the network that has information about any single or set of subscriptions. The complexity of designing this type of systems usually lies on the way to elect the nodes that act as brokers because of the decentralized nature of a distributed network.

In our research \cite{FUTURECOMPUTING14DRW}, we assume that a node can be a publisher, a subscriber, a broker or a combination of these three possibilities. We also assume that all the nodes in the network are able to participate in it without the requirement to adopt the specific role of publisher or subscriber. Nodes that are actively participating in the network but do not take any specific role will be considered as part of the overlay layer. Those nodes of the overlay layer that are able to redirect messages will be considered as brokers.

Event-based systems are classified as topic-based or content-based \cite{Muhl:2006:DES:1162246}. Topic-based systems take into account the subject of messages in order to match publications with subscriptions. Content-based systems use filters to specify the value of subscriptions attributes to redirect notifications. A filter is a boolean function that depends on the set of subscriptions. In our proposal, we plan to deal with a content-based system that uses Bloom filters at broker nodes in order to save memory resources and speed up routing decisions.

Sensor networks frequently use tiny devices with limited battery capabilities that make unsuitable the use of a Global Positioning System (GPS) to disseminate information according to the coordinates of nodes.  In addition to this, the use of virtual coordinates to substitute real coordinates requires the use of sinks or landmarks to structure the network. For these reasons, the use of coordinates in an unstructured sensing scenario is not recommended. We assume that we work in an unstructured scenario in which no routing protocol provides communication between the nodes of the network. 

The constraints of the network infrastructure lead us to the design of a network architecture for distributed event-based systems that must use as less resources as possible (i.e., battery, memory, etc.). In this paper, we present a solution that avoids implying all the nodes of the network in the dissemination process by using a distributed notification service defined by Directional Random Walks (DRWs).

The rest of this paper is organized as follows: Section \ref{sec:state} analyzes the state of the art. Section \ref{sec:methodology} points out the approach to solve the problem specified in this section. Section \ref{sec:research} presents the research efforts already done for the approach specified in Section \ref{sec:methodology}. Section \ref{sec:design} details the process of construction of the proposed architecture. Section \ref{sec:evaluation} evaluates the performance of our solution using DRWs, comparing it with the use of Pure Random Walks (PRWs). Finally, Section \ref{sec:conclusion} summarizes our proposal.

\vspace{0.2em}
\section{State of the art}
\label{sec:state}

\subsection{Distributed and Structured Event-based Systems}

\begin{figure}[!h]
\centering
\includegraphics[width=2.45in]{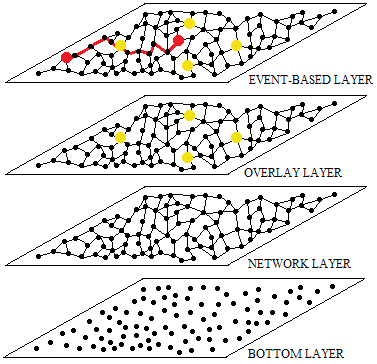}
\caption{Decomposition in layers of the typical design of a distributed and structured event-based system.}
\vspace{-1em}
\label{3layersdistributedusualsystems}
\end{figure}

Distributed and structured event-based systems use three layers on the top of a bottom layer (see Figure~\ref{3layersdistributedusualsystems}), which provides data link functionalities, to facilitate topology control:
\begin{enumerate}
\item The network layer is in charge of providing data forwarding between the different nodes involved in the network. A network protocol, such as the Multicast Ad-hoc On-demand Distance Vector (MAODV) \cite{Roy05securingmaodv:} is needed to provide point-to-point communication.
\item The medium layer is called the overlay layer. It is a virtual layer that builds the event notification service by providing a network of brokers that redirect notifications to the corresponding subscribers. 
\item Finally, on the top layer the event-based protocol is implemented.
\end{enumerate}

One strategy to construct the overlay layer is to use a tree. In TinyMQ \cite{shi2011tinymq}, which is designed specifically for wireless sensor networks, a multi-tree overlay layer is maintained. 

Another strategy is to clusterize the network and use cluster heads to manage messages as in Mires \cite{souto2006mires}, which is a middleware for sensor networks. The Gradient Landmark-Based Distributed Routing (GLIDER) \cite{Fang05glider:gradient} organizes the network using some defined landmarks to compute the Delaunay graph for network partition. Then, the Landmark-Based Information Brokerage scheme (LBIB) \cite{Fang06landmark-basedinformation} uses an overlay layer based in GLIDER to match publishers with subscribers. 

A typical solution is to build the overlay layer using Distributed Hash Tables (DHTs). In these systems, a key is mapped to a particular node with storage location properties. In some DHT architectures, rendez-vous nodes depend on the node ID as in Pastry \cite{Pastry}. In others, as the Content Addressable Network (CAN) \cite{CAN}, a region of the space is used to map a key. Some efforts have been made to apply this solution to sensor networks \cite{Fersi:2013:DHT:2429525.2429572}. When coordinates are available, sensor networks use Geographic Hash Tables (GHT) instead of a typical DHT. Currently, technology companies as Ericsson Research, are making an effort to develop applications that use GHTs in wireless sensor networks \cite{SENSORNETS14M3}.

\subsection{Distributed and Unstructured Event-based Systems}

The main characteristics of distributed and unstructured event-based systems is that they do not maintain an overlay layer. This fact makes easier to deal with network changes. The distributed notification service may be built using flooding, gossiping or random walks.

Most of the algorithms proposed deal with the unstructured nature of wireless communications using flooding to build a tree. A typical solution is to use the On-Demand Multicast Routing Protocol (ODMRP) \cite{Lee:2002}, which is based on the forwarding group concept. Groups are constructed and maintained periodically when a multicast source has data to send. This task is done by broadcasting the entire network with membership information. An extension for ODMRP has been proposed \cite{Yoneki:2004} to adapt a content-based system by adding subscriptions to Bloom filters. Trees also may be configured to self-repair themselves in base to brokers dynamicity \cite{Mottola:2008}. These solutions are reliable but increase the traffic of the network because they use flooding at some point.

Flooding may also be used to continuously exchange subscription information clusterizing the network \cite{Voulgaris06}. Then, notifications are sent to the appropriate cluster, improving the efficiency of the network. Other mechanisms can be used as the combination of a DHT and random walks \cite{Tian:2005}. Cluster heads manage the DHTs while random walks help to connect the different cluster heads of the network. The cluster concept in the network of brokers can be improved in a dynamic scenario by enriching the topology management with predictions based on location \cite{Abdennadher:2013:APM:2508222.2508234}. 

\subsubsection{Probabilistic approaches}
Probabilistic approaches are suitable to deal with dynamicity but they do not offer reliability. Some solutions propose that all the nodes in the network implement a broker that forwards messages to neighbors depending on the estimation of potential subscribers \cite{Haillot:2008}. Other solutions \cite{1437119}, propose to flood subscriptions in a small area and then use random walks to reach these areas. In Quasar \cite{Wong:2008}, subscriptions of a certain area are able to attract or reject notifications, that are propagated with a random walk, using an attenuated Bloom filter \cite{5751342}. A probabilistic solution that uses a random walk specifically designed to go deep into the network is CoQUOS (Continous Queries on Unstructured OverlayS) \cite{Ramaswamy:2011}. Continuous queries are launched to the network using random walks. Peers compute the overlap between their neighbor lists and use this information to forward the random walk to avoid remaining in a cluster. Then, some peers register the query with a probability that depends on the number of hops.







\vspace{0.5em}
\section{Network architecture}
\label{sec:methodology}

Due to the unstructured nature of our network, we propose the development of a dissemination algorithm that merges the network and the overlay layers of a typical distributed and structured event-based system (see Figure~\ref{3layersdistributedusualsystems}). This means that no other network protocol is needed. The main advantage of not using another network protocol is that there is no necessity of maintaining a network topology. This implies that most nodes of the network, which do not actively participate in the process of dissemination, do not have to keep any information about topology. The main consequence is that nodes not involved in the system are able to save energy and computing resources. 

\begin{figure}[!h]
\centering
\includegraphics[width=2.5in]{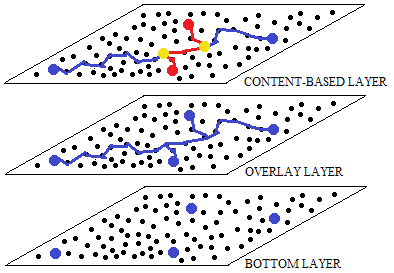}
\caption{Decomposition of the architecture of our design in layers.}
\vspace{-1.00em}
\label{3layers}
\end{figure}

Our design (see Figure~\ref{3layers}) uses two layers on the top of a bottom layer that provides data link functionalities:

\begin{enumerate}
\item The overlay layer is in charge of providing the distributed network of brokers and, at the same time, provides point-to-point communication between publishers and subscribers. The main objective of this strategy is to avoid the use of global information of the network, which is costly to get and maintain.
\item As in Figure~\ref{3layersdistributedusualsystems}, the event-based protocol is implemented at the top layer.
\end{enumerate}

As Section~\ref{sec:introduction} mentions, we assume that a node can be a publisher, a subscriber, a broker or a combination of these three possibilities. Moreover, our design takes advantage of some nodes in the network that want to collaborate. Nodes that participate in the system are considered as part of the overlay layer (blue path of Figure~\ref{3layers}). The overlay layer is formed by the intersection of different publishers and subscribers (blue nodes). Publishers and subscribers implement a DRW until intersecting other DRW. Broker nodes (yellow nodes) are the meeting point between two DRWs.

\begin{figure}[!h]
\centering
\includegraphics[width=2.5in]{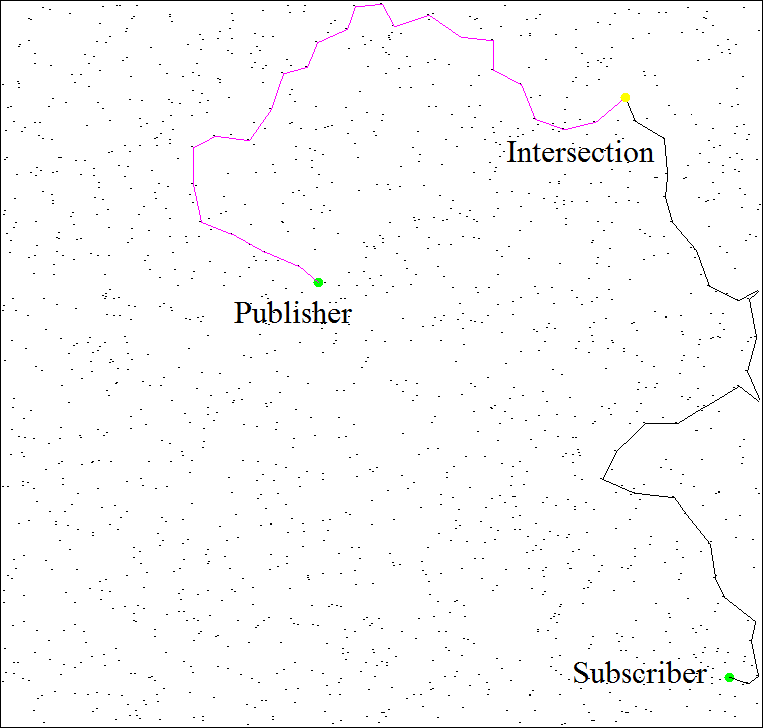}
\caption{Directional random walks intersecting using a Java simulator.}
\vspace{-1em}
\label{sim}
\end{figure}

A DRW is a probabilistic technique able to go forward into the network following a loop-free path. The principle assumed in this strategy is that two lines in a plane cross (see Figure~\ref{sim}). It is unclear how to construct a straight path of relaying nodes in ubiquitous unstructured networks without requiring global information and without making use of geo-coordinates. In this research, two different methods have been proposed in order to build DRWs \cite{ASCOMS13DRW}\cite{SENSORNETS14DRW}.

The strategy followed by the DRW is based on a tabu search \cite{gendreau2014tabu}. The tabu search is a technique used when difficult optimization problems arise. Unfortunately, the theoretical aspects related to a tabu search are so complicated that there is no formal proof of the convergence of the algorithm.

By definition, a tabu search is an iterative procedure in which the next solution is defined by the current solution and a tabu list. A tabu list is a memory that keeps information about the previous iterations of the algorithm. It is used to select the optimal solution for the next iteration. In neighborhood search methods, the tabu list is referred to the set of neighbors of the actual solution. The design of a DRW uses a technique, which is similar to a tabu list based on a neighborhood search method.  A DRW marks the closest nodes of nodes that are already part of the DRW. Afterwards, this information is used to go forward when adding more nodes to the DRW.

The general algorithm for a tabu search based on a neighborhood method is the following:

\vspace{0.2cm}
\begin{tabular}{p{0.43\textwidth}p{0.0\textwidth}}
\begin{itemize}
\item[Step 1] Set initial solution $S_t$ where $t = 0$. Add $S_t$ to the tabu list.
\item[Step 2] Update the current number of iteration $t = t+1$.
\item[Step 3] Create the solution neighborhood $N(S_{t})$ discarding nodes that are part of the tabu list.
\item[Step 4] If $N(S_{t})=\emptyset$ then consider $S_{t-1}$ and return to Step 3.
\item[Step 5] If $N(S_{t})\neq \emptyset$ then evaluate the cost function for all $N(S_{t})$.
\item[Step 6] Select the best solution $S_{t+1}$ basing the choice on the minimum cost. Add $S_{t+1}$ to the tabu list.
\item[Step 7] Stop the algorithm if the stopping criterion is satisfied. Otherwise, return to Step 2.
\end{itemize}
\end{tabular}
\vspace{0.2cm}

A tabu list needs a stopping condition. In our design, the condition is referred to an intersection with a node that is already part of another DRW in the network.



The matching of publishers and subscribers will be done using a special architecture of Bloom filters \cite{5751342} implemented at broker nodes. It is remarkable to mention that in our event-based system no advertisement table is required because filters only manage information about subscriptions. 

Bloom filters are probabilistic data structures that efficiently manage membership of a certain number of elements. The content related to membership is hashed using different hashing algorithms. Then, the positions of the Bloom structure corresponding to the hashes are set to one. The maximum number of elements to be inserted to the filter is fixed in order to maintain a certain probability of false positives. When searching for elements of a certain membership, the corresponding positions of the data structure are checked. The main advantage of Bloom filters is that they do not require much memory space and processing resources; so its use is very convenient in a sensing scenario in which devices have limited capabilities.

In this research, we concentrate on the study of the properties of the overlay layer proposed. It is out of the scope of this work to study a more efficient architecture of Bloom filters at broker nodes for matching publications with subscriptions.

\vspace{0.5em}
\section{Background}
\label{sec:research}


In this section, we present the efforts already made in order to build DRWs.

Based on \cite{ASCOMS13DRW}, a first method to build DRWs is proposed. It is based on the addition of different nodes to the DRW by pre-computing different weights at each node that take into account the two hops path. A weight is defined as follows:
\begin{equation}
\label{DRW Pierre}
n_{xz}^y = \vert \:  N(x)\cap N(z) \: \vert
\end{equation}
where $y$ is the last node added to the DRW; $x$ is the penultimate node added to the DRW; $z$ can be any node of the set $N(y)$ and $N(a)$ is the set of neighbor nodes of node $a$. Furthermore, in this method, a penalty is added to the weight when a node is added to the DRW.

Some properties about our heuristics were found using extensive simulations. The first property claims that DRWs decrease the time to intersection compared to pure random walks. The second property states that cooperation also decreases the time to intersection. Cooperation refers to synchronicity between publishers and subscribers. Finally, it is shown that DRWs are good at balancing the load of the network.

Based on \cite{SENSORNETS14DRW}, a second method to build DRWs is proposed. The main difference with the first design presented for DRWs is that nodes of the first and second neighborhoods of nodes added to the DRW are marked. In addition to this, the cost is not pre-computed, but it is computed when selecting a node as follows:
\begin{equation}
\label{DRW Cristina}
c(v) = \alpha \vert N(v) \cap N(DRW) \vert + \beta \vert N(v) \cap N^2(DRW) \vert
\end{equation}
where $\alpha$ and $\beta$ are parameters used as weights; $v$ can be any node of the set related to the neighborhood of the last node added to the DRW; $DRW$ is the set of nodes that are part of the DRW; $N(a)$ is the set of neighbor nodes of node $a$ and $N^2(DRW)$ is the set of neighbor nodes of $N(DRW)$.

In the first part of this research, the properties associated with a DRW were assessed. Implementations of DRWs of one or two branches were studied. The main results show that the use of one branch is as efficient as the use of two branches. Moreover, it is shown that the use of second neighborhoods to forward the DRW does not improve the Euclidean distance traversed in the network. It is also shown that shorter paths are obtained when using higher densities of nodes in the network. In the second part of this research, an information brokerage system was evaluated using a double ruling method. As in the first paper, it is shown that the algorithm is efficient at balancing the load using a few nodes of the network. In fact, we can state that the method proposed is as good as a traditional Rumor Routing algorithm \cite{Braginsky:2002:RRA:570738.570742} with an infinite memory.

\vspace{0.5em}
\section{Design of the overlay layer}
\label{sec:design}

\subsection{Network Model}
A DRW is defined in a graph $G=(V, E)$, where $V$ is the set of vertices and $E$ is the set of edges. $u, v \in V$ are connected $u \sim v$ if $(u, v) \in E$. The size of $G$ is denoted by $\mid V \mid = n$ and the number of edges is denoted by  $\mid E \mid = m$. We denote the neighborhood of $v$ as $N(v)=\{ u \in V \hspace{0.1cm} | \hspace{0.1cm} u \sim v\}$.

\subsection{Implementation of the overlay layer}

In order to assess the architecture proposed, we have used a variation of the algorithm presented at \cite{SENSORNETS14DRW}. 

The set of edges and vertices associated to a DRW of ID $x$ are denoted by $E'_x$ and $V'_x$. Our technique consists of selecting the set of vertices $V'_x$ that are part of the DRW. Each vertex of $V'_x$ is denoted by $v'_{x, i}$. The current number of nodes in the active path of the DRW is denoted by $i$. Vertices are chosen consecutively until two DRWs intersect.

A vertex $v$ is selected to be part of the DRW as $v'_{x,i}$ if it has the minimum cost at iteration $i$ between $N(v'_{x, i-1})$. The cost function used is a varitation of the cost function used in (\ref{DRW Cristina}):
\begin{equation}
\label{Cost}
c(v) = \vert N(v) \cap N(DRW_{x, i}) \vert 
\end{equation}
where $N(DRW_{x, i})$ denotes the set of neighbors of $V'_x$ at iteration $i$. Formally, it is defined as:
\begin{equation}
\label{DRW first neighborhood}
N(DRW_{x, i}) =  \left [\bigcup_{j=0}^{i} N(v'_{x, j}) \right ]
\end{equation}

The use of $N(DRW_{x, i})$ is of particular interest to our research because it allows us to exploit the broadcast advantage of the wireless medium. This process can be seen as a repulsion mechanism to force a branch to keep moving forward. The result of this mechanism is that neighbors that are not part of $N(DRW_{x, i})$ have higher possibilities to be added to the DRW.

\begin{figure}[!h]
\centering
\includegraphics[width=\linewidth]{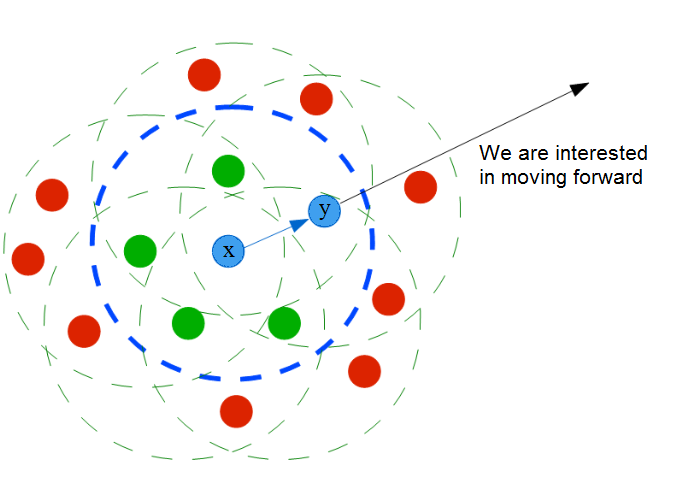} \\
\footnotesize a) Two nodes added to the directional random walk \vspace{1em} \\
\vspace{2.02em}
\includegraphics[width=\linewidth]{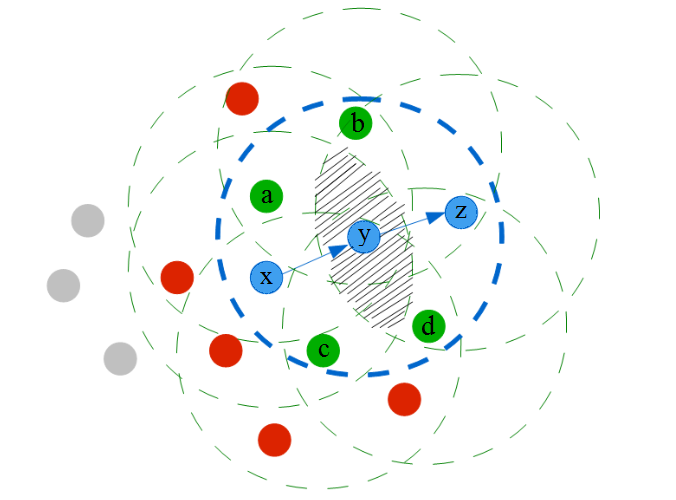} \\
\footnotesize b) Three nodes added to the directional random walk
\vspace{0.22em}
\caption{Construction of a directional random walk.}
\vspace{-3.02em}
\label{ConstructionofaDRW}
\end{figure}

Figure \ref{ConstructionofaDRW} illustrates the selection of a node $z$ when nodes $x$ and $y$ have already been added to a DRW. Before adding $z$ to the DRW (see Figure \ref{ConstructionofaDRW}.a), neighbors of the penultimate node added to the DRW are marked as part of the neighborhood of the walker. At this stage, node $y$ has to select the next node to be added to the DRW between its neighbors $a$, $b$, $c$, $d$ and $z$ (see Figure \ref{ConstructionofaDRW}.b). In order to avoid remaining in the same zone, we are interested in selecting a candidate that helps to push the DRW to other zones of the network. Candidate $a$ has a cost of 3 because it has three neighbors marked as part of the neighborhood of the walker (see Figure \ref{ConstructionofaDRW}.b). Candidates $b$, $c$, $d$, and $z$; have a cost of 2, 3, 2 and 1, respectively. Candidate $z$ has the minimum cost, so that it is added to the DRW and neighbors of node $y$ are marked as part of the neighborhood of the walker.

Publishers and subscribers are considered as initiators of a DRW. Algorithm~\ref{alg:overlay} shows how DRWs are intersected for a certain number of publishers and subscribers in the network. 
 
Algorithm~\ref{alg:drw} shows the pseudocode used for the construction of a DRW. The selection of a node is based on the computation of a cost (line 26). A candidate node is added to the DRW if it has the minimum cost between all the candidate nodes (line 30). A node is considered as candidate, if it is part of the neighborhood of the last node added to the branch (line 21). It is assumed that there are no candidate nodes when the neighborhood of the last node added is empty or all of them are already part of the DRW (line 18). In order to assure intersections our algorithm goes back in the DRW to search for the nearest non traversed neighbor.

\begin{algorithm}
\begin{algorithmic}[1]
\REQUIRE $Number\hspace{0.1cm} of\hspace{0.1cm} total\hspace{0.1cm} initiators: I$

\STATE Define Thread $drw0$ as a new \textsc{DRW($v'_{0, 1}$)}
\STATE Define Thread $drw1$ as a new \textsc{DRW($v'_{1, 1}$)}

\STATE Start Thread $drw0$
\STATE Start Thread $drw1$

\STATE \textbf{while}\{(Intersection value of $drw0$ is 0) $\&\&$ (Intersection value of $drw1$ is 0)\}
\STATE \textbf{end while}

\FOR {$i=2; i < I; i++$}
	\STATE Define Thread $drwi$ as a new \textsc{DRW($v'_{i, 1}$)}
	\STATE Start Thread $drwi$
	\STATE \textbf{while}\{(Intersection value of $drwi$ is 0)\} 
	\STATE \textbf{end while}
\ENDFOR

\end{algorithmic}
\caption{Construction of the overlay layer}
\label{alg:overlay}
\end{algorithm}

\begin{algorithm}
\begin{algorithmic}[1]
\REQUIRE $Initiator: v'_{x, 1} \in V$
\STATE { \textbf{function} \textsc{DRW($v'_{x, 1}$)} }
\STATE $Intersection=0$
\STATE {\bfseries add} $v'_{x, 1}$ to $V'_x$
\IF{any $v \in N(v'_{x, 1})$ is part of other DRW}
	\STATE {\bfseries return} $Intersection=1$
\ELSE
	\IF{any $v \in N(v'_{x, 1})$ is part of other DRW}
		\STATE $v'_{x, 2} \gets v$
		\STATE {\bfseries return} $Intersection=1$
	\ELSE
		\STATE select $v'_{x, 2} \in N(v'_{x, 1})$ randomly
		\STATE $i=2;$
		\WHILE{$Intersection=0$}
			\FOR { each $v \in N(v'_{x, i-1})$}
				\STATE {\bfseries add} $v$ to $N(v'_{x, i})$
			\ENDFOR
			\STATE $i++;$
			\WHILE{$(\{v \hspace{0.1cm} | \hspace{0.1cm} v \in N(v'_{x, i})\} = \emptyset) \hspace{0.1cm}||$ \\ $\hspace{0.1cm} (\{v \hspace{0.1cm} | \hspace{0.1cm} v \in N(v'_{x, i}) \} \in DRW)$}
				\STATE $i--;$					
			\ENDWHILE
			\FOR {each $v \in N(v'_{x, i})$}
				\IF{$v$ is part of other DRW}
					\STATE $v'_{x, i} \gets v$
					\STATE {\bfseries return} $Intersection=1$
				\ELSE
					\STATE compute $c(v)$ as defined by (\ref{Cost})
				\ENDIF					
			\ENDFOR
			\IF{$Intersection=0$}	
				\STATE {\bfseries add} to the DRW $v$ $\hspace{0.1cm} | \hspace{0.1cm} v, min\{c(v)\}\in  N(v'_{x, i})$
			\ENDIF
		\ENDWHILE
		\STATE {\bfseries return} $Intersection=1$
	\ENDIF		
\ENDIF
\STATE {\bfseries end function}
\end{algorithmic}
\caption{Thread of the construction of a DRW}
\label{alg:drw}
\end{algorithm}

\vspace{0.5em}
\section{Evaluation of the novel architecture}
\label{sec:evaluation}
To assess the performance of the overlay layer we have implemented a Java simulator. The  networks used for the numerical evaluation have been obtained by placing nodes randomly and uniformly in a squared area of side size $1\times1$ with a range of communication of $r = 0.05$. The communication model is defined by the range of communication. Two nodes that are closer than the range of communication can communicate. The graph we obtain in this way is often referred by Unit Disc Graph (UDG). Under these conditions, it is hard to obtain connected networks with less than 1.000 nodes, so we have conducted numerical validation for more dense networks assuring that they are completely connected. The total number of nodes considered has been 1.000, 2.000 and 3.000.

\begin{figure}[!h]
\centering
\includegraphics[width=2.5in]{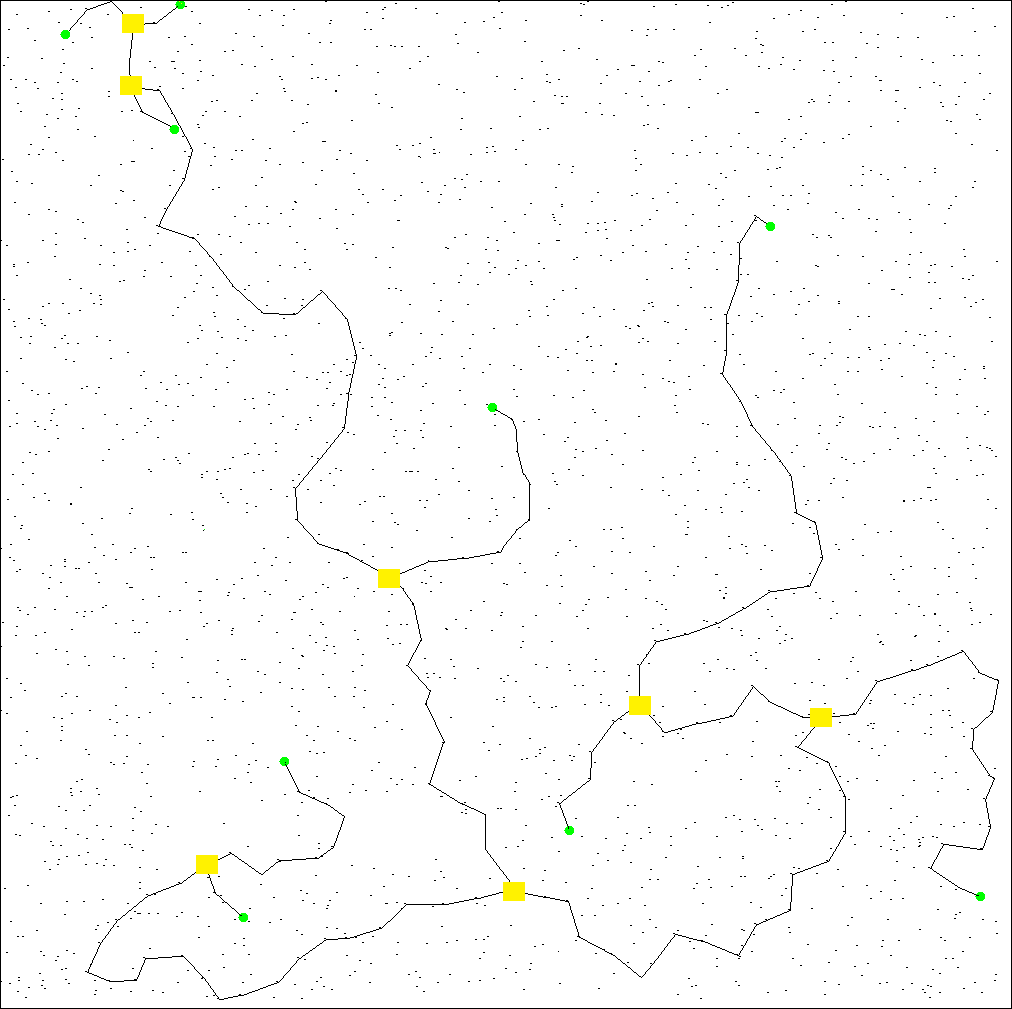}
\caption{Overlay layer using directional random walks in a Java simulator.}
\vspace{-1.02em}
\label{overlaylayer}
\end{figure}

As previously mentioned, for the implementation of the overlay layer we use different DRWs that intersect. Figure~\ref{overlaylayer} shows a simulation of the overlay layer in which yellow squares represent the distributed network of brokers while publishers and subscribers are represented using green circles.

\begin{figure}[!h]
\centering
\includegraphics[width=2.5in]{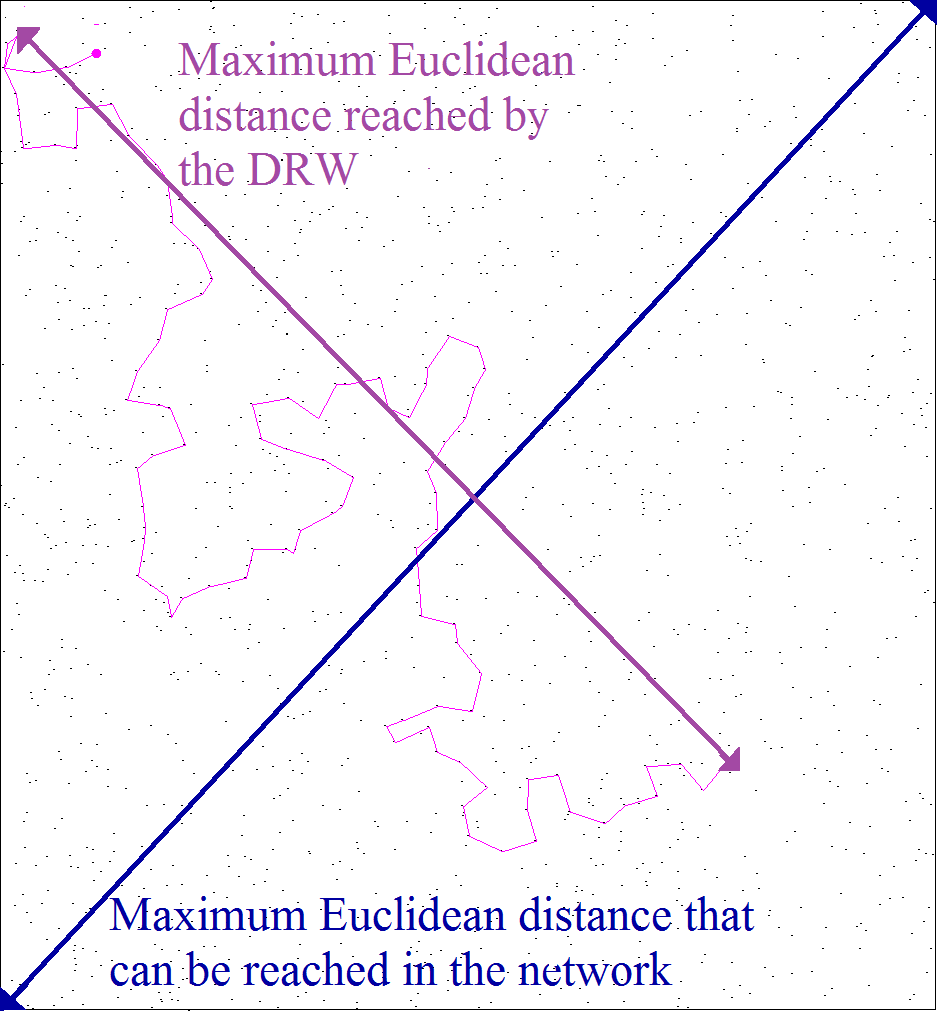}
\caption{Representation of the $depth$.}
\vspace{-1.02em}
\label{depth}
\end{figure}

The performance metric used is the $depth$ (see Figure \ref{depth} and (\ref{eq:depth})). The $depth$ compares the maximum Euclidean distance reached by all the nodes that are part of the list of relaying nodes of any DRW, with the maximum Euclidean distance that can be reached in the network. It is defined as:

\begin{equation}
depth(Overlay\hspace{0.1cm}Layer) = \frac{max\{\{d(v'_i, v'_j) \hspace{0.1cm} | \hspace{0.1cm} v'_i, v'_j \in  \bigcup_{x}V'_x\}}{max\{d(v_i, v_j) \hspace{0.1cm} | \hspace{0.1cm} v_i, v_j \in V\}}
\label{eq:depth}
\end{equation}
where $d$ denotes the Euclidean distance.

The evaluation of the design of the overlay layer is based on: 
\begin{itemize}
\item The number of nodes in the active path.
\item The $depth$.
\end{itemize}

For this purpose, we assess the performance of the architecture proposed for different number of publishers and subscribers in the network. Our results, have been obtained by using extensive simulations for each network considered. Box and whisker plots are used to visualize data. As previously mentioned, we considered networks of 1.000, 2.000 and 3.000 nodes. The total number of networks simulated totals 126.000.

The total number of simulations when evaluating networks of 1.000 nodes has been 20.000. We have evaluated the performance of the overlay layer for the following total number of publishers and subscribers: 2, 3, 4, 5, 6, 7, 8, 9, 10, 20, 30, 40, 50, 75, 100, 250, 500, 625, 750 and 875. For each total number of initiators, we have simulated 100 different networks. 

Similarly, for networks of 2.000 nodes we have obtained 21.000 simulations using a total number of initiators of: 2, 3, 4, 5, 6, 7, 8, 9, 10, 20, 30, 40, 50, 75, 100, 250, 500, 1.000, 1.250, 1.500 and 1.750. 

Finally, for networks of 3.000 nodes 22.000 simulations have been conducted for a total number of initiators of: 2, 3, 4, 5, 6, 7, 8, 9, 10, 20, 30, 40, 50, 75, 100, 250, 500, 1.000, 1.500, 1.875, 2.250 and 2.625.

Moreover, we have used PRWs for comparison with our method, which select the next node of the walker randomly. So that overlay layers formed by PRW have also been evaluated for 1.000, 2.000 and 3.000 nodes.

\subsection{Impact on the number of nodes in the active path}

\begin{figure*}[!t]
\centering{
\begin{tabular}{cc}
 	 \includegraphics[width=0.5\linewidth]{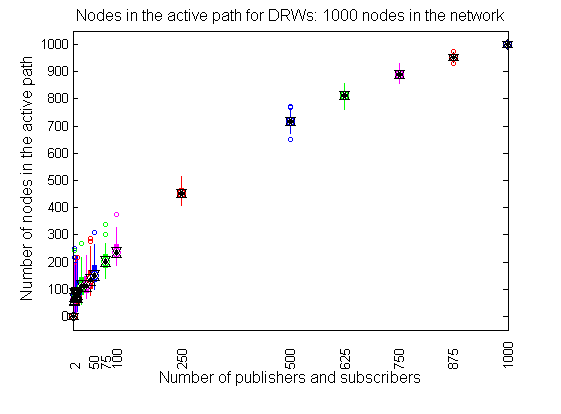}
 	    \label{fig:label:a} &
 	 \includegraphics[width=0.5\linewidth]{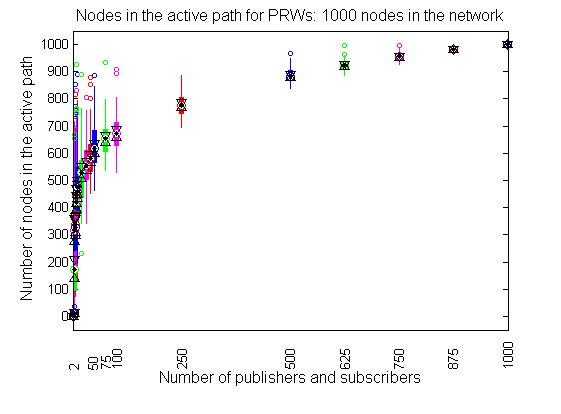}%
		\label{fig:VT falses positives _2objectsAND}\\ 
  	 \scriptsize a) & \scriptsize b)\\ 
     \includegraphics[width=0.5\linewidth]{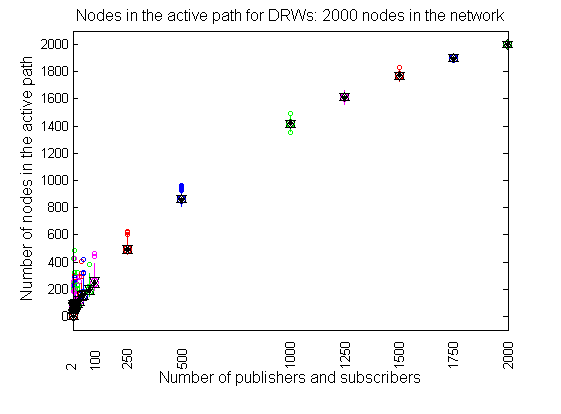}%
		\label{fig:Gain ORs} &
     \includegraphics[width=0.5\linewidth]{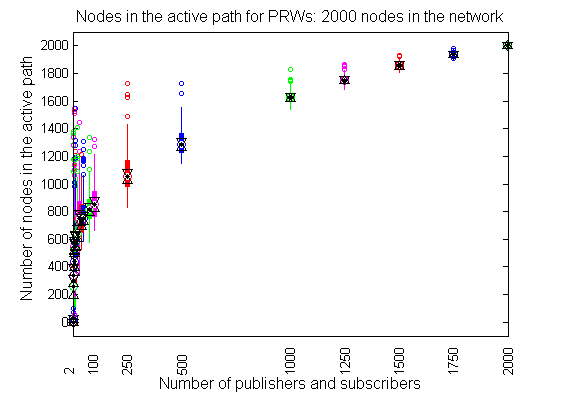}%
		\label{fig:Gain _3objectsAND}\\ 
     \scriptsize c) & \scriptsize d)\\ 
     \includegraphics[width=0.5\linewidth]{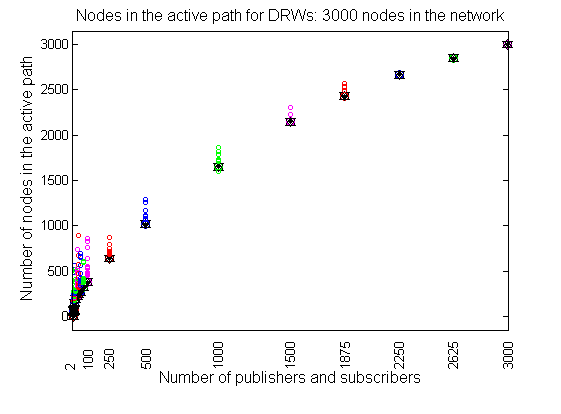}%
		\label{fig:Gain ORs} &
     \includegraphics[width=0.5\linewidth]{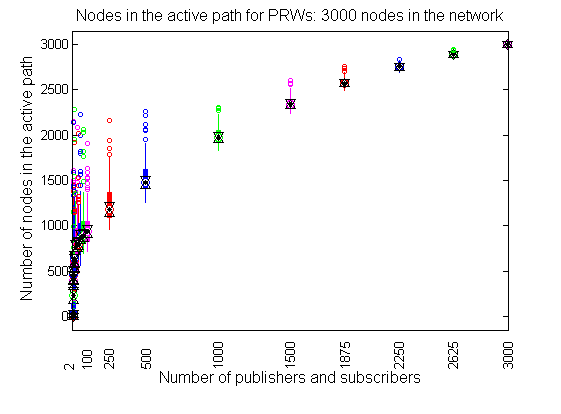}%
		\label{fig:Gain _3objectsAND}\\ 
     \scriptsize e) & \scriptsize f)
\end{tabular}}
\centering
\caption{Nodes in the active path of the overlay layer for different number of publishers/subscribers in the network using directional random walks (a, c, e) and pure random walks (b, d, f). The total number of nodes in the network is 1.000 nodes (a, b), 2.000 nodes (c, d) and 3.000 nodes (e,f).}
\label{fig:activepath}
\end{figure*}

Figure \ref{fig:activepath} compares the results obtained by using DRWs and PRWs for networks of 1.000, 2.000 and 3.000 nodes.

Interestingly, Figures \ref{fig:activepath}.a, \ref{fig:activepath}.c and \ref{fig:activepath}.e, that show the performance of the overlay layer when using DRWs are proportionally, almost identical for different densities of nodes in the network following the same logarithmic behavior. This behavior suggests that the base of the logarithm decreases as the density of the network is increased. Nevertheless, Figures \ref{fig:activepath}.b, \ref{fig:activepath}.d and \ref{fig:activepath}.f, that show the performance of the overlay layer when using PRWs, present a different logarithmic behavior. We observe that proportionally, less dense networks grow faster in terms of nodes in the active path; but still, the more nodes we have in the network, the more nodes we have in the active path. This means that as in the previous case, the base of the logarithm decreases as the density of the network is increased; but the change on the base is more dramatic for DRWs. This effect is produced because DRWs reduce the random component attached to the experiment, allowing to have a more predictable performance, that is traduced in more directionality  or similarity to a straight line of the walker.

Besides this, overlay layers constructed using PRWs present more outliers that overlay layers that use DRWs. The result is that DRWs present a more reliable performance in the construction of the active path of our architecture. In addition to this, it is obvious, that overlay layers that use PRWs present larger active paths.

As previously mentioned, in all cases the more nodes we have in the network, the more nodes we have in the active path of the overlay layer for the same number of publishers and subscribers in the network. This consequence is reasonable, because the less density of nodes we have in the network, the less candidate nodes we have to construct the walker. The result of this performance is that less dense networks are saturated before.

\subsection{Impact on the depth}

\begin{figure*}[!t]
\centering{
\begin{tabular}{cc}
 	 \includegraphics[width=0.5\linewidth]{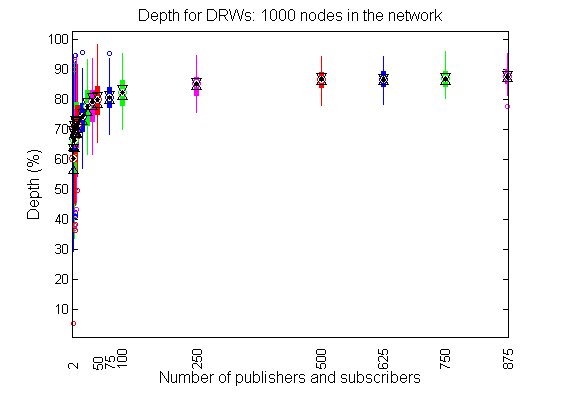}
 	    \label{fig:label:a} &
 	 \includegraphics[width=0.5\linewidth]{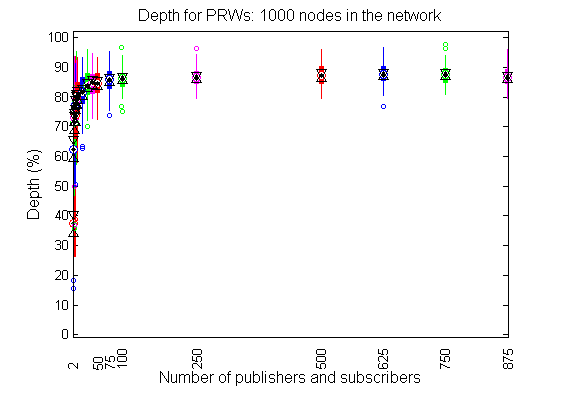}%
		\label{fig:VT falses positives _2objectsAND}\\ 
  	 \scriptsize a) & \scriptsize b)\\ 
     \includegraphics[width=0.5\linewidth]{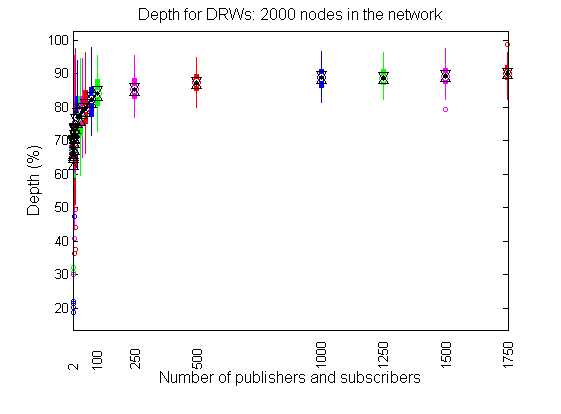}%
		\label{fig:Gain ORs} &
     \includegraphics[width=0.5\linewidth]{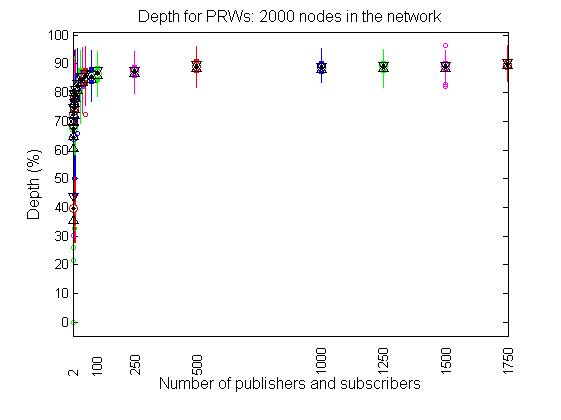}%
		\label{fig:Gain _3objectsAND}\\ 
     \scriptsize c) & \scriptsize d)\\ 
     \includegraphics[width=0.5\linewidth]{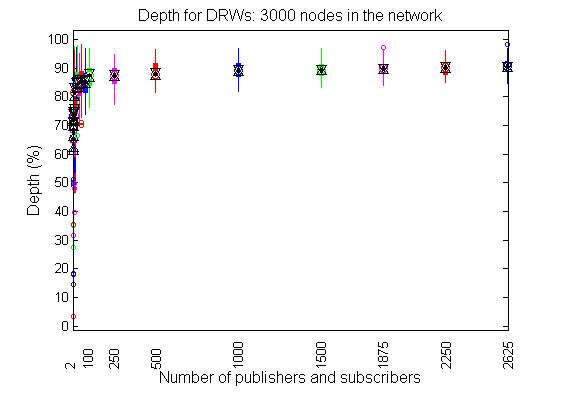}%
		\label{fig:Gain ORs} &
     \includegraphics[width=0.5\linewidth]{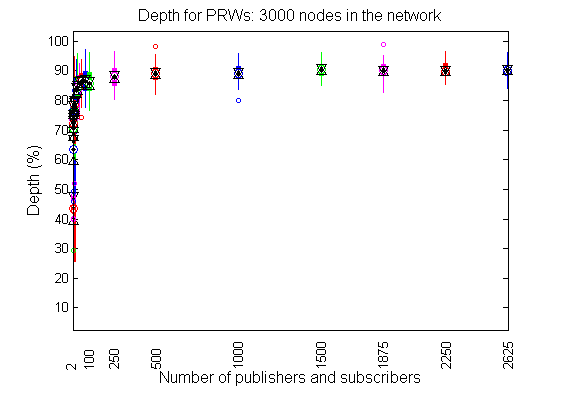}%
		\label{fig:Gain _3objectsAND}\\ 
     \scriptsize e) & \scriptsize f)
\end{tabular}}
\centering
\caption{Depth of the overlay layer for different number of publishers/subscribers in the network using directional random walks (a, c, e) and pure random walks (b, d, f). The total number of nodes in the network is 1.000 nodes (a, b), 2.000 nodes (c, d) and 3.000 nodes (e,f).}
\label{fig:depth}
\end{figure*}

Figure \ref{fig:depth} shows the resulting $depth$ for the different densities of networks considered using DRWs and PRWs. The main conclusion extracted is that $depths$ are very similar for all cases and that the maximum Euclidean distance that is going to be traversed in the network is reached very early.

\begin{figure*}[!t]
\centering{
\begin{tabular}{cc}
 	 \includegraphics[width=0.5\linewidth]{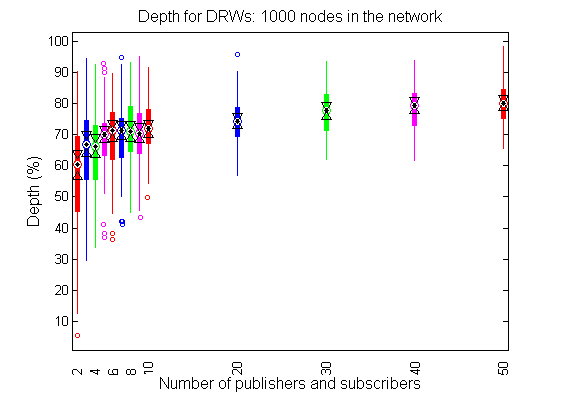}
 	    \label{fig:label:a} &
 	 \includegraphics[width=0.5\linewidth]{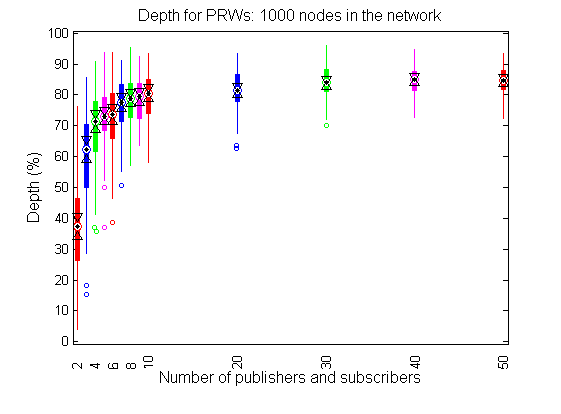}%
		\label{fig:VT falses positives _2objectsAND}\\ 
  	 \scriptsize a) & \scriptsize b)\\ 
     \includegraphics[width=0.5\linewidth]{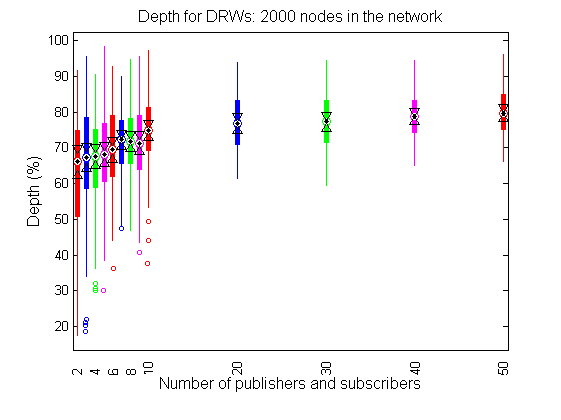}%
		\label{fig:Gain ORs} &
     \includegraphics[width=0.5\linewidth]{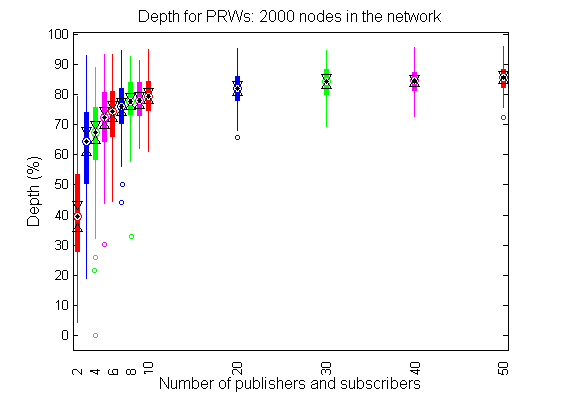}%
		\label{fig:Gain _3objectsAND}\\ 
     \scriptsize c) & \scriptsize d)\\ 
     \includegraphics[width=0.5\linewidth]{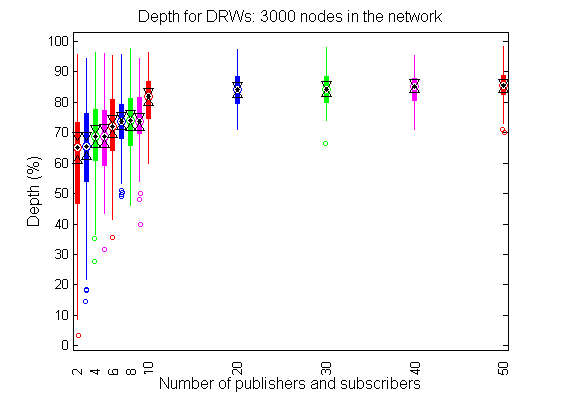}%
		\label{fig:Gain ORs} &
     \includegraphics[width=0.5\linewidth]{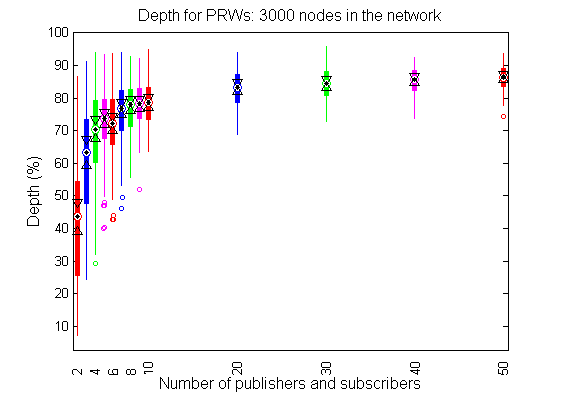}%
		\label{fig:Gain _3objectsAND}\\ 
     \scriptsize e) & \scriptsize f)
\end{tabular}}
\centering
\caption{Depth of the overlay layer, for the first nodes in the active path, for different number of publishers/subscribers in the network using directional random walks (a, c, e) and pure random walks (b, d, f). The total number of nodes in the network is 1.000 nodes (a, b), 2.000 nodes (c, d) and 3.000 nodes (e,f).}
\label{fig:depth_50}
\end{figure*}

Figure \ref{fig:depth_50} shows in detail the performance when having a few number of publishers and subscribers in the network. In all cases, the depth is importantly increased when having three publishers and subscribers in the network. Furthermore, we observe that the directionality of DRWs leads to increase the $depth$ when having two publishers or subscribers compared to the $depth$ reached by PRWs.

Finally, we can state that there exists no correlation between the number of nodes in the active path and the $depth$. So that other factors, as the density of the network, have more impact in the number of nodes of the active path.

\vspace{0.5em}
\section{Conclusion}
\label{sec:conclusion}

In this paper, a  novel network architecture for distributed event-based systems that use sensing devices has been proposed. We present a network architecture that merges the network and overlay layers of typical structured event-based systems. Our results, validated through extensive simulations, show that DRWs are suitable for the construction of an overlay layer that provides point-to-point communication and a distributed notification service. 

Our strategy avoids using other network protocol to provide point-to-point communication. This implies that most nodes of the network, which do not actively participate in the process of dissemination, do not have to maintain any information about topology. The main consequence is that nodes not involved in the system are able to save energy and computing resources.

We evaluate the performance of the overlay layer using DRWs and PRWs for its construction. Our results show that for our purpose DRWs are more efficient than PRWs. This is due, mainly to the good properties of DRWs, which use less nodes of the network for the establishment of the active path of the overlay layer. Moreover, we can state that overlay layers that use DRWs have a more reliable performance that overlay layers that use PRWs. Furthermore, it is remarkable to mention that, in all cases, the more nodes we have in the network, the more nodes we have in the active path of the overlay layer for the same number of publishers and subscribers in the network. Finally, it is interesting to discard any correlation between the number of nodes that form the overlay layer and the maximum Euclidean distance that is traversed by the walkers; mainly because the maximum $depth$ is quickly reached.

\vspace{0.27em}
\section*{Acknowledgment}

This work has been developed as part of the POPWiN project that is financially supported by the Swiss Hasler Foundation in its “SmartWorld - Information and Communication Technology for a Better World 2020” program.

\bibliographystyle{IEEEtran}
\bibliography{ref}

\end{document}